\documentstyle{article}
\input boxedeps.tex
\SetRokickiEPSFSpecial  
\HideDisplacementBoxes
\def \F{\phi}

\def\r{\rho}

\def\a{\alpha}

\def\s{\sigma}

\def\half{{1\over 2}}
\def\d{\dagger}
\def\be{\begin{equation}}
\def\eq{\end{equation}}
\def\Tr{{\rm Tr}}

\def\q2{{\rm QCD}_{2}}
\def\q4{{\rm QCD}_{4}}

\begin{document}
\begin{flushright}
CERN-TH/97-87\\
hep-th/9704211
\end{flushright}
\begin{center}
{\Large Collinear QCD Models}\footnote{Invited talk at 
`New Non-Perturbative Methods and  Quantisation on the Light-Cone', 
Les Houches, France, 24 Feb - 7 Mar 1997. (To appear in the proceedings.)}\\
\vspace{15mm}
{\bf S. Dalley}\footnote{On leave from: Department of Applied
Mathematics and Theoretical Physics, Cambridge University,
Silver Street, Cambridge CB3 9EW, England.}\\
\vspace{5mm}
{\em Theory Division, CERN, CH-1211 Geneva 23, Switzerland}
\vspace{5mm}
\end{center}
In ancient times,
't Hooft studied the mesons in $QCD_{1+1}$ \cite{hoof} to illustrate the
power of the large $N$ limit \cite{hoof2} in the light-cone formalism. 
Recently some generalisations of the 't Hooft model have been studied,
which retain a remnant of transverse degrees of freedom, based on a
dimensional reduction of QCD to $1+1$ dimensions 
\cite{col1,col2,col3,col4,col5,col6,col7}.
In this collinear approximation, quarks and gluons are artificially
restricted to move in one space dimension, but retain their
polarization degree of freedom.
In this lecture, a problem which in principle involves a large number
of partons will be addressed in the context of the collinear model at large
$N$. 
For light-cone quantisation, large numbers of partons are synonymous
with small Bjorken-$x$. 
The example treated here\footnote{Another example was treated in the lecture
--- asymptotic spectrum of glueballs ---  for
which there was not enough room in these proceedings.}
is the quark distribution function in a heavy meson, which is
supposed to exhibit a version of Regge behaviour at small-$x$.
The central idea  involves high light-cone 
energy boundary conditions on wavefunctions --- ladder relations ---
which typically connect Fock space sectors of differing numbers of
partons. The same ideas carry over to $3+1$ dimensions
\cite{abd}.

We start from $SU(N)$ gauge theory in $3+1$-dimensions 
with one flavour of quarks.
If we pick an arbitrary fixed space direction $x^3$ and
restrict ourselves to zero momentum in the transverse directions
\be
\partial_{x_{\perp}} A_{\mu} = \partial_{x_{\perp}} \Psi  = 0 \label{tzero}
\eq
for the gauge and quark fields,
one finds an
effectively two-dimensional gauge theory of adjoint scalars and 
fundamental Dirac spinors with action 
\begin{eqnarray}
S & = & \int dx^0 dx^3 \  -{1 \over 4g^2} \Tr\{F_{a b}F^{ab}\}
+ \Tr\{ -\half \bar{D}_{a} 
\F_{\r} \bar{D}^{a} \F^{\r} \nonumber \\
&& -{tg^2 \over 4} [\F_{\r},\F_{\s}][\F^{\r},\F^{\s}]  
+ \half m_{0}^{2} \F_{\r}\F^{\r} \}  
+{{\rm i} \over \sqrt{2}} (\bar{u} \gamma^{a}
D_{a} u + \bar{v} \gamma^{a} D_{a} v)  \\
&& + {m_F \over \sqrt{2}} (\bar{u} v
+ \bar{v} u)  
 - {sg\over \sqrt{2}} (\bar{u} (\F_1 + {\rm
i}\gamma^{5} \F_2)
u - \bar{v}(\F_1 - {\rm i} \gamma^{5} \F_2 )v) \nonumber \ ,\label{red}
\end{eqnarray}
where $a$ and $b \in \{ 0,3\}$, $\r \in \{ 1,2 \}$,
$\gamma^{0} = \s^1$, $\gamma^{3} = {\rm i} \s^2$,
$\gamma^{5} = {\rm i} \s^1 \s^2$, 
$\F_{\r} = A_{\r}/g$, 
$\bar{D}_{a} = \partial_{a} + {\rm i}[A_{a},.]$, $D_{a} =
\partial_{a} + {\rm i}A_{a}$, 
$\int dx^1 dx^2 = {\cal L}^2$, $g^2 = \tilde{g}^2 / {\cal L}^2$, and 
$\tilde{g}$ is the four-dimensional coupling.
The two-component
spinors $u$ and $v$ are related to $\Psi$ by 
\begin{eqnarray}
 \Psi  = {1 \over 2^{1/4} {\cal L}} \left( \begin{array}{c} 
 u_{R+} \\  u_{L+} \\  u_{L-} \\  u_{R-}  \end{array} \right)
\ , \
 u = \left( \begin{array}{c} 
u_{L+} \\ u_{R+} \end{array} \right) \ , \ v = \left( \begin{array}{c}
u_{L-} \\ u_{R-} \end{array} \right) \ .
\label{spin}
\end{eqnarray}
The suffices $L$ $(R)$ and $+$ $(-)$ in (\ref{spin}) 
represent Left (Right) movers and
$+ve$ $(-ve)$ helicity, which is a conserved quantity.
Thus $u,v,\F_1 ,\F_2$ represent the transverse polarisations of the
$3+1$
dimensional quarks and gluons. Since the dimensional reduction
procedure
treats space asymmetrically, dimensionless parameters
$s$ and $t$, and a bare gluon mass $m_0$ can occur due to loss of
transverse local gauge transformations. 
For the present 
application the precise choice of $s$ and $t$  will not be 
qualitatively important, so we set $s=1$ and $t=0$ for simplicity.

In the light-cone  gauge $A_- = (A_0 - A_3)/\sqrt{2} = 0$, the fields 
$A_+$ and $u_{L \pm}$ are non-propagating in light-front time $x^+ =
(x^0 + x^3)/\sqrt{2}$ so may be eliminated by their constraint equations
\be
\partial_{-} u_{L\pm}  =  {\rm i} F_{\mp} \ \ , \ \ 
(\partial_{-})^2 A_+  =   g^2 J^+  \label{const1} 
\eq
\begin{eqnarray}
       F_{\pm} & = &  {m_F \over \sqrt{2}} u_{R\pm} \pm gB_{\mp} u_{R\mp} 
\label{eff} \\
J^+ &  = &  {\rm i} [B_-, \partial_{-} B_+ ] + {\rm i} [B_+,
\partial_{-} B_- ]  + u_{R+}u^{\dagger}_{R+} + u_{R-}u^{\dagger}_{R-} 
\end{eqnarray}         
and $B_{\pm} = (\F_1 \pm {\rm i} \F_2)/\sqrt{2}$.
The exchange of non-propagating particles associated with the
constrained fields results non-local interactions
in the light-cone hamiltonian
\begin{equation}
P^- =  \int d x^-  \ \
     F_+^{\dagger} \frac{1}{{\rm i}\partial_-} F_+ + 
     F_-^{\dagger} \frac{1}{{\rm i}\partial_-} F_-  
  -\frac{g^2}{2} \Tr \left\{ J^+ \frac{1}{(\partial_-)^2} J^+ \right\}
+ \frac{m_{0}^{2}}{2}\Tr \ \F^2
      \label{ham}
\end{equation}
The zero momentum limit of the constraints (\ref{const1})
forces a  condition
on the quark-gluon combined system
\be
\int_{-\infty}^{+\infty} dx^- F_{\pm} = 0 \ \ , \ \ 
\int_{-\infty}^{+\infty} dx^- J^{+} = 0 \label{zero} \ ,
\eq
assuming $u_L$ and $\partial_{-} 
A_+$ vanish at $x^{-} = \pm \infty$. This is required
for finiteness of the non-local interactions  in (\ref{ham}). 
The relation involving $J^+$, which amongst other things restricts one
to gauge singlets, will not be discussed further here.

Introducing the harmonic 
oscillator modes
of the physical fields\footnote{The superscript on $k^+$ has been dropped
for clarity; $i,j \in \{1,\ldots,N\}$ are gauge indices and
$\dagger$ is now understood as the quantum complex conjugate, so
does not transpose them.}
\begin{eqnarray}
u_{R\pm i} & =& {1 \over \sqrt{2\pi}} \int_{0}^{\infty} dk
\left( b_{\pm i}(k) {\rm e}^{-ikx^-} 
+ d^{\d}_{\mp i}(k) {\rm
e}^{ikx^-} \right)  \label{mode1}\\
B_{\pm ij} & =& {1 \over \sqrt{2\pi}} \int_{0}^{\infty} {dk \over 
\sqrt{2k}} \left( a_{\mp ij}(k) {\rm e}^{-ikx^-} + a_{\pm
ji}^{\d}(k) {\rm e}^{ikx^-} \right) \label{mode2}
\end{eqnarray}
in the quantum theory we can expand any hadron state $|\Psi(P^+)>$ 
of total momentum $P^+$ 
in terms of a Fock basis \cite{lepage}.
The operators $a_{\pm}^{\dagger}$ create gluons with helicity $\pm 1$,
while $b_{\pm}^{\dagger}$ and $d_{\pm}^{\dagger}$ 
correspond to quarks and antiquarks
(respectively) with helicities $\pm \frac{1}{2}$.
At large $N$ a gauge-singlet meson is  a superposition 
\begin{eqnarray}
|\Psi (P^+)>  & =&  \sum_{n=2}^{\infty} \int_{0}^{P^+} dk_1 \dots
dk_n \sum_{\alpha_i = \pm}
   \hspace{1mm} \delta(k_1 + \cdots + k_n - P^+) \times   \nonumber \\
&&  \frac{ f_{\alpha_1 \dots \alpha_n}
( k_1,  k_2, \dots,  k_n)}{\sqrt{N^{n-1}}}    \times  \\
 && d^{\dagger}_{\alpha_1 i }( k_1) a^{\dagger}_{\alpha_2 ij}
( k_2) a^{\dagger}_{\alpha_3 jk}( k_3) \dots
a^{\dagger}_{\alpha_{n-1} lm}( k_{n-1}) 
b^{\dagger}_{\alpha_n m}( k_n) |0> \nonumber
\label{mesonbs}
\end{eqnarray}
Introducing the fourier transform $\tilde{F}(w)$,
one finds that in the quantum theory Eq.(\ref{zero}) 
can be meaningfully applied as an annihilator of
physical states for the cases
\begin{eqnarray}
 \lim_{w \to 0^+}    {\tilde F}_{\pm i}
 (w) \cdot |\Psi(P^+)> & = & 0 \label{quark} \\
 \lim_{w \to 0^-} {\tilde F}_{\pm i}^{\dagger}
(w) \cdot |\Psi(P^+)> & = & 0 \label{aquark} \ .
\end{eqnarray}
The first relation yields a condition on the Fockspace wavefunctions $f$ 
involving vanishing quark momentum $k=w>0$, the second on vanishing anti-quark
momentum $k=-w>0$: 
\begin{eqnarray}
\lefteqn{ f_{\mp \pm \alpha_1 \cdots \alpha_n}
(k , k_1,\dots, k_{n+1})} & & \nonumber  \\ 
& = & \pm \lambda \left[
 \frac{f_{\pm \alpha_1 \cdots \alpha_n}
  (k  + k_1,  k_2, \dots,  k_{n+1})}
    {\sqrt{k_1}} \right. \nonumber \\
& & + \int_0^{\infty}\frac{dp dq}{\sqrt{q}} 
\hspace{1mm} \delta (p + q - k) f_{\pm \mp \pm\alpha_1 \cdots \alpha_n}
( p,  q, k_1, \dots ,  k_{n+1}) \left.  \frac{}{}  \right] \ , \label{first}
\end{eqnarray}
\begin{eqnarray}
\lefteqn{ f_{\mp \mp \alpha_1 \cdots \alpha_n}
( k , k_1,\dots,  k_{n+1})} & & \nonumber  \\ 
& = & \pm \lambda
\int_0^{\infty}\frac{dp dq}{\sqrt{q}} 
\hspace{1mm} \delta (p + q - k) f_{\pm \mp \mp \alpha_1 \cdots \alpha_n}
( p,  q, k_1, \dots , k_{n+1}) \ ,
\label{second}
\end{eqnarray}
with $\lambda = \sqrt{g^2 N /2 \pi m^{2}_{F}}$ and a similar set of
relations for quarks from (\ref{quark}); in (\ref{first})(\ref{second})
the limit $k \to 0^+$ is understood.
If we adopt the following momentum-space 
operator ordering in $P^-$
(\ref{ham})
\be
 \int_{-\infty}^{0} {dw \over w} 
  {\tilde F}_{\pm i}^{\dagger}     {\tilde F}_{\pm i}  - 
\int_{0}^{\infty} {dw \over w} {\tilde F}_{\pm i} {\tilde
F}_{\pm i}^{\dagger}
 \ , \label{form}
\eq
we then apparently have manifest finiteness as $w \to 0$ for physical states.
Normal ordering the oscillator modes in  $P^-$ would spoil finiteness. 
Since we do
 not  normal order the form (\ref{form}), 
infinite quark self energies (self-inertias)
are generated but  no vacuum energies 
are generated. 

However the above argument is flawed by the fact that
infinities may also arise due to integration over the parton
momenta in the wavefunction $|\Psi>$, since (\ref{first})(\ref{second})
are to be interpreted at {\em fixed} $k_i$ as $k \to 0^+$.
This is evident from the light-cone Schrodinger equation, obtained by 
projecting $2P^+ P^- |\Psi> = {\cal
M}^2 |\Psi>$  onto a specific $n$-parton Fock state
\be
\left(\left[{{\cal M}^2\over 2P^+}\right] 
- \sum_{i=1}^{n} {m_i^2 \over2 p_i}\right)  f_{\alpha_1 \dots \alpha_n} 
(p_1, \dots, p_n) = \hat{V}\left[f_{\alpha_1 \dots \alpha_n} 
( p_1,  \dots,  p_n)\right]
\label{full}
\eq
where $\hat{V}$ is the interaction kernel (including self-inertias), 
${\cal M}$ the
boundstate mass, and $m_i$ is $m_{F}$ (quark)
or $m_B$ (gluon). The ladder relations (\ref{first})(\ref{second})
are necessary for finiteness of the internal integrations in $\hat{V}$
at fixed external momenta $p_i$.
However further renormalisation  of $\hat{V}$  is
necessary since the ladder relations do not ensure  finiteness when one or
more external momenta $p_i$ vanish in (\ref{full}).\footnote{This 
point, and also the
integral terms in (\ref{first})(\ref{second}),
were missed in ref.\cite{col5}.}
In fact an explicit two-loop calculation 
of the fermion self-energy in light-cone ${\rm Yukawa}_{1+1}$ \cite{perry}
shows that divergences do not cancel for the same cutoff
on all small momenta. In general the renormalisation that cures these
divergences will depend on the precise cut-off(s) employed. It has been 
suggested to renormalise the fermion kinetic mass finitely to restore
parity invariance in light-cone calculations \cite{burk}, and
this should coincide with ensuring finiteness of ${\cal M}$.\footnote{After
this lecture was typed, a preprint appeared \cite{burk2} 
which verifies this for certain examples with a Yukawa interaction
only.}

Eqs.(\ref{first})(\ref{second}) show that
 the meson wavefunction components do not vanish as the
quark momentum vanishes. It will be demonstrated that this leads directly
to a rising quark distribution function at small $x= k/P^+$.
The probability to find an anti-quark --- the answer is the same for a quark
--- with momentum fraction $x= k/P^+$ of the meson is
\begin{eqnarray}
Q(x)  & = &  \sum_{n=2}^{\infty} \sum_{\alpha_i}
 \int_0^{P^+} dk_1 \dots dk_n 
   \hspace{1mm} \delta(k_1 + \cdots + k_n - P^+) \times \nonumber \\  
& &  \delta (k_1 - k) \hspace{1mm}
 |f_{\alpha_1 \dots \alpha_n}
(k_1, k_2, \dots,  k_n)|^2 \nonumber \\
&  = & \sum_{\a} <d^{\d}_{\a i} d_{\a i} (x)> 
\label{qxsf}
\end{eqnarray} 
For the polarized version $\Delta Q(x)$ one inserts ${\rm sgn}(\a)$.
In order to make use of (\ref{first})(\ref{second}) to evaluate
$Q(x \to 0)$, it is helpful to eliminate the integral terms, which generate
renormalisation of the other (non-integral) terms.
Although the integrals are over a set of measure zero, they are
non-zero due to the singular behaviour of the integrand.
This singular behaviour can be found from the (correctly renormalised)
Schrodinger equation (\ref{full}).
Let us consider the helicity $+1$ meson with valence component $f_{++}$.
The general idea is to use an expansion in $\lambda$ and $\log{1/x}$ to 
evaluate (\ref{qxsf}). The leading orders we shall calculate
are independent of any additional fermion kinetic mass renormalisation
in  (\ref{full}).
The leading log approximation amounts to considering
the integration region $ k_{n-2}  >> k_{n-2} >> \cdots >> k_2 >> k_1=k$
in (\ref{qxsf}).
In this region we may use the ladder relations iteratively to express
every $n$-parton non-valence contribution in terms of $f_{++}$. For example,
truncating to no more than one gluon 
we obtain $Q(x \to 0) = \lambda^2 < 1/y >_{++}$
from the $f_{-++}$ component, where $< 1/y >_{++}$ is the average inverse
momentum fraction in the $f_{++}$ component. 

Truncating to no more than two gluons, we 
can compute some of the subleading $\lambda$ and $\log{1/x}$ corrections.
For this we need to use the $n=4$ boundstate equation (\ref{full}) 
(for which there
is neither a fermion self-inertia nor a finite kinetic mass renormalisation) 
to evaluate the integral
term in (\ref{first}). The following results neglect
the $A_+$ exchange process
between quark and gluon in (\ref{full}), whose effects cannot be explicitly
resummed. From resumming instantaneous fermion ($u_{L}$) processes  
the correct ladder relations in this 2-gluon approximation become
\begin{eqnarray}
f_{-++}(x \to 0, x_1, x_2) &  =  & \lambda^* {f_{++}(x_1, x_2) 
\over \sqrt{x_1}} \\
f_{+-++}(x \to 0, x_1, x_2, x_3) & = & - \lambda {f_{-++}(x_1, x_2, x_3) 
\over \sqrt{x_1}} 
\end{eqnarray}
where
\be
\lambda^* = { \lambda \over 1 + \lambda^2 \left[ { \log{(m_{F}^{2}/m_{B}^{2})}
\over 1- (m_{B}^{2}/m_{F}^{2})} \right]}
\eq
Then to leading log the contributions from one and two gluon components of
the wavefunction give
\be
Q(x \to 0) \approx (\lambda^{*})^2 (1 + \lambda^2 \log{1/x}) < 1/y >_{++} \ .
\eq
An example of  a  next-to-leading log contribution comes from 
integrating the two-gluon contribution over the
region $\int_{0}^{k} dk_2 \int_{k} dk_3$ using (\ref{full}) at $n=4$
\be
\lambda^2 (\lambda^*)^2 
\log{\left(1 + {m_{F}^{2} \over 
m_{B}^{2}} \right) } < 1/y >_{++} 
\eq
\begin{figure}
\centering
\BoxedEPSF{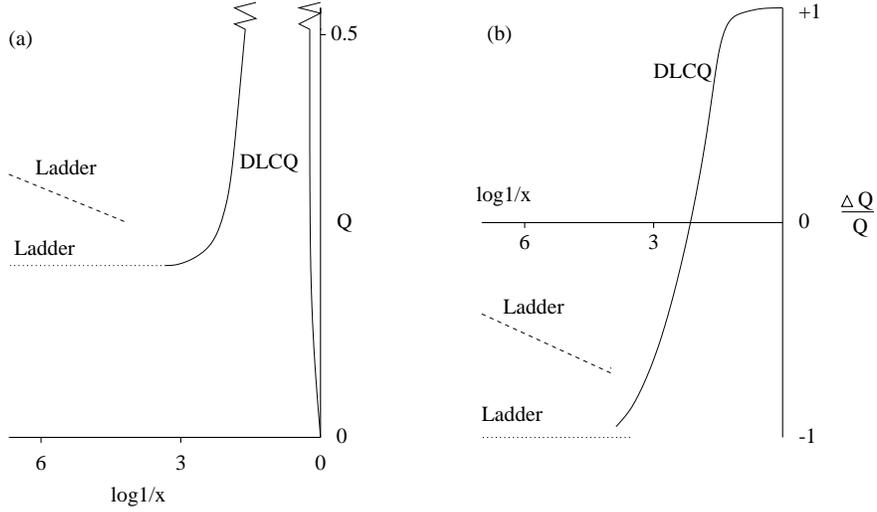 scaled 450}
\caption{(a) Unpolarized distribution function for helicity
$+1$ meson at $\lambda^2 = 0.1$: (Dotted) 1-gluon ladder prediction
($ < 1/y >_{++}$ is indistinguishable from 2 for heavy quarks);
(solid) DLCQ up to 1 gluon, K=24; (dashed) arbitrary-gluon 
tree-level ladder prediction.(b) helicity asymmetry $\Delta Q / Q$.}

\label{Fig1}
\end{figure}
\begin{figure}
\centering
\BoxedEPSF{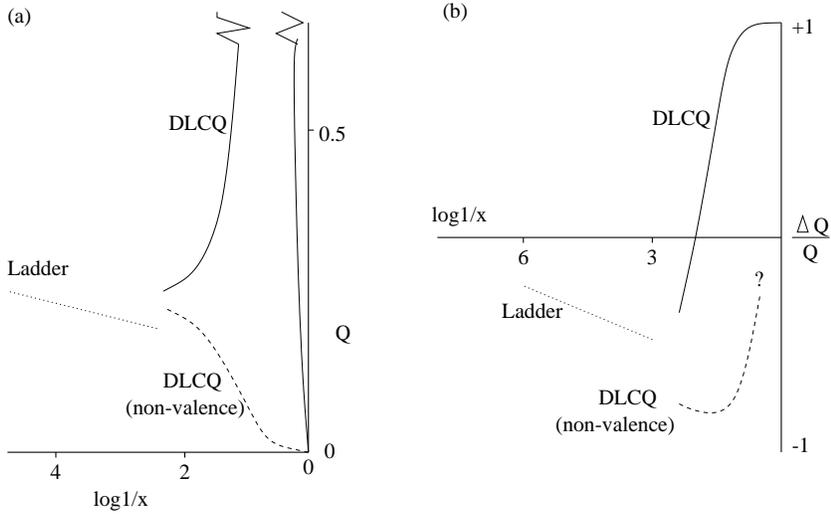 scaled 450}
\caption{(a) Unpolarized distribution function for helicity
$+1$ meson at $\lambda^2 = 0.1$, $m_{B}^{2}/m_{F}^{2} = 0.1$: 
(Dotted) ladder prediction for up to 2 gluons and next-to-leading log;
(solid) DLCQ up to 2 gluons, K=15; (dashed) non-valence part of the
DLCQ calculation. (b) Helicity asymmetry $\Delta Q / Q$.}
\label{Fig2}
\end{figure}

The analytic ladder results are compared with a non-perturbative 
DLCQ solution  of (\ref{full}) truncated to the same number of gluons
in Figs. 1 and 2.
The DLCQ calculations are formal at finite $x$ since no additional
fermion kinetic mass renormalisation
has been carried out to ensure finite ${\cal M}$ as $K \to \infty$.
In practice however, for heavy quarks the
effects of this omission are very tiny at finite $K$, 
and the plots shown should be a good
representation of the exact result at finite $x$ (the same comment applies to
plots in ref.\cite{col5}).  At $x \sim 1/K$ the DLCQ
results should in any case match onto the analytic predictions.
There are many sources in the approximations which might account
for the remaining
discrepancy in the normalisation and slope seen in Fig.2 at small $x$, 
although the agreement is much better if the DLCQ calculation is repeated
without the quark-gluon $A_+$ exchange.

More generally if we consider
the ladder relations to leading order in $\lambda$, which means neglecting
the integrals in (\ref{first})(\ref{second}), 
an arbitrary number of gluons can be eliminated
to yield an exponential sum of leading $\log{1/x}$'s (Fig.1)
\be 
Q(x \to 0) \approx \lambda^2 x^{-\lambda^2} < 1/y >_{++} 
\eq
The integral terms
should only renormalise $\lambda$ in the previous expression.
All the results point to a rising small-$x$ unpolarized distribution.
At large $N$ the polarization 
asymmetry changes sign and then vanishes at small
$x$ in the leading log approximation, $\Delta Q / Q \sim -x^{2 \lambda^2}$.
\vspace{5mm}

Acknowledgements: I have had many useful discussions about collinear
models with F. Antonuccio, S. Brodsky, M. Burkardt, H-C. Pauli
and B. van de Sande. The DLCQ computer program is due to
F. Antonuccio \cite{thesis}. I thank P. Grang\'{e} for
co-ordinating a stimulating workshop.

\vfil

\begin{thebibliography}{100}
\bibitem{hoof}  't Hooft G.,  Nucl. Phys.,  B75, 1974, 461.
\bibitem{hoof2}  't Hooft G.,  Nucl. Phys.,  B72, 1974, 461.
\bibitem{col1} Dalley S., Klebanov I. R.,  Phys. Rev.,
  D47, 1993, 2517.
\bibitem{col2} Demeterfi K., Klebanov I. R., Bhanot G., 
Nucl. Phys.,  B418, 1994, 15.
\bibitem{col3} Pauli H. C., Kalloniatis A. C., Pinsky S., 
Phys. Rev.,  D52, 1995, 1176.
\bibitem{col4} Antonuccio F., Dalley S.,  Nucl. Phys.,
  B461, 1996, 275. 
\bibitem{col5} Antonuccio F., Dalley S.,  Phys. Lett.,  B376, 1996, 154.
\bibitem{col6} Pauli H. C., Bayer R.,  Phys. Rev., D53, 1996, 939.
\bibitem{col7} van de Sande B., Burkardt M.,  Phys. Rev., D53, 1996, 4628. 
\bibitem{abd} Antonuccio F., Brodsky S. J., Dalley S., preprint CERN-TH/97-88.
\bibitem{lepage} Lepage G. P., Brodsky S. J.,  Phys. Rev.,
D22, 1980, 2157.
\bibitem{brod} Pauli H. C., Brodsky S. J.,   Phys. Rev.,
 D32, 1985, 1993.
\bibitem{perry} Harindranath A., Perry R.,  Phys. Rev.,
  D43, 1991, 4051.
\bibitem{burk} Burkardt M., Phys. Rev., D54, 1996, 2913.
\bibitem{burk2} Burkardt M., preprint [hep-th/9704162].
\bibitem{thesis} Antonuccio F., D.Phil Thesis, University of Oxford (1996).
\end{thebibliography}
\end{document}